\documentclass[11pt,a4paper,twocolumn]{article}

\usepackage[T1]{fontenc}
\usepackage[utf8]{inputenc}
\usepackage[margin=0.75in]{geometry}
\usepackage{amsmath}
\usepackage{amssymb}
\usepackage{amsthm}
\usepackage{graphicx}
\usepackage[expansion=false,protrusion=true]{microtype}
\usepackage[hidelinks]{hyperref}
\usepackage[compact]{titlesec}
\usepackage{enumitem}
\usepackage{etoolbox}

\titlespacing*{\section}{0pt}{1.4ex plus 0.4ex minus 0.3ex}{0.8ex plus 0.2ex}
\titlespacing*{\subsection}{0pt}{1.0ex plus 0.3ex minus 0.2ex}{0.5ex plus 0.1ex}
\titlespacing*{\paragraph}{0pt}{0.9ex plus 0.2ex minus 0.2ex}{0.6em}
\setlist{itemsep=1pt,parsep=1pt,topsep=2pt,partopsep=0pt,leftmargin=1.6em}

\newcommand{\Prob}{\mathbb{P}}
\newcommand{\Qm}{\mathbb{Q}}
\newcommand{\Exp}{\mathbb{E}}

\title{Kinetic Interference in Translational Control:\\
A Path-Measure Framework for Collision-Triggered Transcript Decay}

\author{Shlomo Segal\\
\small Independent Theoretical Researcher}

\date{}

\begin{document}

\maketitle

\begin{abstract}
\noindent
I connect two literatures that have developed independently: the path-measure formulation of
non-equilibrium statistical mechanics, in which a trajectory action decomposes into a
time-antisymmetric (entropic) and a time-symmetric (frenetic) sector, and the stochastic
modelling of ribosomal traffic on messenger RNA. The biological target is a proposed
intervention---antisense oligonucleotide (ASO) interference with wobble-uridine
($\mathrm{U}_{34}$) modification of transfer RNA---whose intended effect is not to abolish
translation but to perturb its \emph{timing}, driving ribosome collisions and
collision-triggered transcript decay preferentially on high-flux, codon-biased transcripts.

Dynamical-activity and large-deviation analyses of generic lattice exclusion models---most
notably the symmetric and totally asymmetric simple exclusion processes---are well established
\cite{lecomte2012,gjw2007,quasteltsai2021}. To my knowledge, while these frameworks exist for
generic exclusion dynamics, their formalization specifically for ribosomal traffic queues,
translation elongation, and collision-triggered no-go decay remains unoccupied; this paper
addresses that narrower gap, not the general one.

Two claims here are load-bearing and untested. First, \emph{selectivity}: that transcripts
whose loss is therapeutically desirable are separable, by vulnerable-codon-pair burden, from
transcripts whose loss is toxic. Second, \emph{non-redundancy}: that the frenetic
decomposition yields predictions, specific to ribosomal queueing and collision-triggered
decay, not already obtainable either from rate-level exclusion-process models or from
existing generic activity/large-deviation analyses of exclusion processes. This paper
establishes neither. It specifies both as falsifiable tests with pre-registered decision
rules, including the outcomes under which the framework should be abandoned or substantially
narrowed. It is a research programme proposal, not a result. No new experimental,
computational, or bioinformatic results are reported.
\end{abstract}

\section{Introduction}

\subsection{Rate versus abundance in translational control}

Most therapeutic modulation of gene expression is \emph{abundance-directed}: reduce the
quantity of a transcript or of the protein it encodes. An alternative axis is
\emph{rate-directed}: leave abundance intact and perturb the velocity at which the existing
machinery operates. In a cell held far from equilibrium, where steady-state concentrations
reflect a balance of fluxes rather than a minimum of free energy, the two are not equivalent.
A perturbation to timing can have consequences that no perturbation to abundance reproduces.

Translation elongation is the natural test case. The elongation cycle is driven, cyclic, and
multi-state; ribosomes are extended objects that exclude one another on a shared
one-dimensional track; and the cell operates dedicated machinery that responds not to slow
translation as such, but to the \emph{collision} of one ribosome with a stalled predecessor.
This is what makes rate perturbation potentially non-linear. A modest increase in dwell time
at a codon does nothing until queue formation crosses a threshold, past which a qualitatively
distinct pathway---collision-triggered decay---engages.

\subsection{Kinetic traps}

I use \emph{kinetic trap} for a perturbation a system cannot adapt to by the mechanisms it
ordinarily uses to resist perturbation. Resistance to an abundance-directed agent is usually
achieved by sequence change---mutation of a binding site, amplification of a target.
Resistance to a rate-directed perturbation acting on the shared decoding apparatus would
instead require reshaping codon usage across many transcripts at once: slower, more
constrained.

\emph{Trap} is, at present, a metaphor, not a theorem, and I will not pretend otherwise. To
earn the term the framework needs a fitness or flux functional, a space of accessible
compensatory states---isodecoder upregulation, modification-enzyme upregulation, codon drift,
reduced initiation rate---and a demonstration that no low-cost escape path exists within it.
Section~\ref{sec:roadmap} lists this as open, not settled.

\subsection{Relation to existing non-equilibrium formalism}
\label{sec:priorart}

None of the mathematical apparatus below is new. Path-measure decompositions into
time-antisymmetric and time-symmetric sectors, and the associated notion of dynamical
activity, are established for Markov jump processes and diffusions in general
\cite{maes2019,basumaes2015}, and have been applied specifically to lattice exclusion
processes: large-deviation functions for entropy production in the totally asymmetric simple
exclusion process (TASEP) are known in closed form \cite{sahamukherji2015}, the full
hydrodynamic large-deviation principle for TASEP has been established
\cite{quasteltsai2021}, and, specifically for the symmetric simple exclusion process (SSEP),
an inactive dynamical phase driven by constraining the time-symmetric activity rather than
the current has been characterized by Lecomte, Garrahan, and van Wijland
\cite{lecomte2012}, building on the broader $s$-ensemble programme for large deviations of
dynamical activity developed by Garrahan, Jack, Lecomte, van Wijland, and co-workers in the
context of kinetically constrained models \cite{gjw2007}. This body of work already does
time-symmetric activity analysis on driven lattice systems, exclusion processes included. It
is not absent from the literature, and I will not describe it that way anywhere in this
programme.

What it does not address is the ribosome-specific structure motivating the present paper:
extended particles running a multi-state internal elongation cycle, sequence-dependent
(codon-level) hopping rates, and---the essential point---a collision-triggered
quality-control response that removes a stalled particle from the lattice, converting the
process into an exclusion system with state-dependent removal rather than a closed lattice
gas. To my knowledge, no published work formalizes the time-symmetric sector of \emph{this}
process, or connects it to collision-triggered no-go decay as a biological readout. My claim
to originality, such as it is, rests entirely on this narrower combination---narrower than
any claim of having introduced time-symmetric path-measure analysis to exclusion processes as
such.

That the frenetic sector can carry information genuinely absent from the entropic sector,
rather than merely re-expressing it, is not something I assume; it has been established
constructively in a different setting. In a reaction-channel-selection problem for overdamped
gradient flows, I have shown that when two histories share both endpoints under a
conservative force the entropy difference between them vanishes identically, so no
dissipation-based criterion---including transition-state and Kramers--Langer theory evaluated
at the saddle---can distinguish the channels even in principle, while the frenetic sector
alone determines the branching ratio \cite{segal2026}. That is an existence proof that
entropic degeneracy with frenetic discrimination is a genuine physical regime, not an
artifact of the decomposition. It is not evidence that such a regime exists for ribosomal
collisions specifically. That is exactly what the non-redundancy requirement of
Section~\ref{sec:redundancy} and Milestone~2 (Section~\ref{sec:milestone2}) still need to
settle.

\subsection{Scope of the present work}

This is a conceptual paper. It proposes a framework, identifies the tests that would validate
or refute it, and states the conditions under which it fails. No experiment, no simulation,
no bioinformatic analysis. Where the underlying biology comes from published work, that work
is in model organisms under genetic, not pharmacological, perturbation; extrapolation to
human cells and to ASO-mediated intervention is hypothesis, not established fact.

\section{The Physics Backbone}
\label{sec:physics}

\subsection{Path measures and the action decomposition}

Consider the translational system over a time window $[0,T]$, described by a trajectory
$\omega$ drawn from a path measure $\Prob$. Relative to a reference measure $\Qm$, define the
action
\begin{equation}
\mathcal{A}[\omega] \;=\; -\log \frac{d\Prob}{d\Qm}[\omega].
\label{eq:action}
\end{equation}
Writing $\Theta$ for time reversal, $\mathcal{A}$ decomposes into antisymmetric and symmetric
parts,
\begin{align}
\mathcal{A}[\omega] &\;=\; -\tfrac{1}{2}\, S[\omega] \;+\; \tfrac{1}{2}\, K[\omega],
\label{eq:decomposition}\\
S[\omega] &\;=\; \mathcal{A}[\Theta\omega] - \mathcal{A}[\omega], \notag\\
K[\omega] &\;=\; \mathcal{A}[\omega] + \mathcal{A}[\Theta\omega], \notag
\end{align}
where $S$ is the entropic sector and $K$ the frenetic sector. For an overdamped diffusion the
Onsager--Machlup form of \eqref{eq:action} makes the split explicit, the antisymmetric part
recovering the entropy production and the symmetric part collecting the dynamical activity;
this overdamped case is the one in which I demonstrated entropic degeneracy and frenetic
discrimination constructively \cite{segal2026}, and it is the setting against which the
Markov-jump construction used below should ultimately be checked for consistency. The general
formalism, its realizations for Markov jump processes and diffusions, and the associated
response theory are established \cite{maes2019,basumaes2015}; I adopt them rather than extend
them.

$S$ tracks net directed flux; $K$ tracks undirected dynamical activity---traffic, escape
rates, residence times. Two systems can carry identical currents while differing in activity,
and the frenetic sector is the object that distinguishes them.

\subsection{The reference-measure problem}
\label{sec:reference}

The entropic sector $S$ is reference-independent across time-reversal-symmetric references;
the frenetic sector $K$ is not. Under a change $\Qm \to \Qm'$, the frenetic term shifts by
the symmetric part of $\log (d\Qm'/d\Qm)$. So "the intervention acts through the frenetic
sector" is, absent further argument, a statement about bookkeeping, not about the system.
This is the principal mathematical vulnerability of the framework, and I address it with two
explicit commitments.

First, fix a canonical reference. For the elongation jump process with rates
$k(x \to y)$, take the geometric-mean symmetrization
\begin{equation}
q(x \to y) \;=\; q(y \to x) \;=\; \sqrt{\,k(x \to y)\, k(y \to x)\,},
\label{eq:reference}
\end{equation}
which is time-reversal-symmetric and preserves the symmetric rate skeleton of the dynamics.

Second---the more important commitment---I do not claim $K$ itself is invariant. I claim only
that the \emph{perturbation-induced difference}
\begin{equation}
\Delta K \;=\; \Exp_{\Prob_{\epsilon}}\!\left[K\right] \;-\; \Exp_{\Prob_{0}}\!\left[K\right]
\label{eq:deltaK}
\end{equation}
is reference-independent \emph{under a stated condition}: that the perturbation acts
multiplicatively on a fixed transition graph, without opening or closing transitions. Where
that condition fails---notably if the intervention introduces a new absorbing channel, as
sequestration of tRNA would---the cancellation between $\Prob_0$ and $\Prob_\epsilon$ fails,
and the residual ambiguity must be bounded explicitly. For admissible references
$\Qm, \Qm'$ that are $\Theta$-invariant and mutually absolutely continuous with both
$\Prob_0$ and $\Prob_\epsilon$ on $[0,T]$, a bound of the form
\begin{equation}
\left| \Delta K^{\Qm} - \Delta K^{\Qm'} \right|
\;\le\;
\left\| \log \tfrac{d\Qm'}{d\Qm} \right\|_{\infty}
\cdot
\left\| \Prob_{\epsilon} - \Prob_{0} \right\|_{\mathrm{TV}}
\label{eq:bound}
\end{equation}
should be reported alongside any central estimate. Whether the proposed intervention satisfies
the multiplicative condition is an open item
(Section~\ref{sec:milestone2}).

\subsection{Dwell times, queueing, and the collision observable}

Ribosomal traffic is conventionally modelled as a totally asymmetric exclusion process (TASEP)
or its mean-field ribosome-flow reductions, with codon-dependent hopping rates, extended
particles occupying $\ell > 1$ successive codons, and an initiation rate $\alpha$ controlling
density \cite{reuveni2011}. Kinetic modelling in this tradition has established that
collision with a trailing ribosome, not slow-down as such, is the event that engages quality
control, and that collisions may serve as a timer for it \cite{subramaniam2017}.

Two cautions. First, collision-triggered rescue makes the service process state-dependent on
the queue itself: a collided ribosome is removed by quality-control machinery, so the right
object is a queue with reneging, not a standard exclusion process with a slow site, and
standard steady-state results do not transfer unmodified. Second, projecting the multi-state
elongation cycle---accommodation, GTP hydrolysis, proofreading rejection, peptidyl transfer,
translocation---onto a scalar reaction coordinate generically breaks the Markov property of
the projected process, and the decomposition \eqref{eq:decomposition} presupposes that it
does not. Any application must either justify the projection by timescale separation or
verify robustness to memory kernels.

\subsection{The non-redundancy question}
\label{sec:redundancy}

I state the objection that most threatens this section against the correct baseline. The
relevant comparison is not rate-level exclusion-process machinery in general, but
specifically the existing dynamical-activity and large-deviation literature for SSEP and
TASEP \cite{lecomte2012,gjw2007,sahamukherji2015,quasteltsai2021}, which already extracts
time-symmetric information---activity fluctuations, inactive dynamical phases,
large-deviation rate functions for entropy production---from exclusion dynamics without
invoking a biological collision-decay readout. Everything the present framework currently
predicts about ribosomal collisions---threshold initiation rates, queue lengths, spacing
dependence, decay flux---is in principle computable from the codon-level rate constants using
existing exclusion-process and queueing machinery, and the qualitative phenomenon of an
activity-driven regime change has a documented precedent in the inactive dynamical phase of
driven exclusion lattices \cite{lecomte2012}. The decomposition \eqref{eq:decomposition} is,
absent further argument, a re-description of quantities already accessible by these routes.
If it yields no experimentally distinguishable prediction beyond them, the physics is
ornamental twice over: redundant with rate-level modelling, and redundant with existing
activity-based treatments of exclusion processes.

The framework owes a specific deliverable, and the bar is higher than in a generic
lattice-gas setting: a regime, arising specifically from the ribosome's multi-state
elongation cycle, its extended and sequence-dependent hopping rates, or its
collision-triggered removal dynamics, in which the frenetic sector predicts collision
statistics that neither a matched rate-level TASEP model \emph{nor} a generic SSEP/TASEP
activity-fluctuation analysis reproduces. I do not have one. Identifying such a regime, or
establishing that none exists---in which case the correct conclusion is that existing
exclusion-process activity theory already suffices and the present framework reduces to an
application rather than an extension---is the second pre-registered milestone
(Section~\ref{sec:milestone2}).

\section{The Biological Hypothesis}

\subsection{The proposed causal chain}

The hypothesis has five links.
\begin{enumerate}
\item An ASO, delivered to the nucleus, sterically occupies a region of the pre-tRNA required
for access by $\mathrm{U}_{34}$-modification machinery (Elongator, ALKBH8, CTU).
\item The resulting mature tRNA population is hypomodified at $\mathrm{U}_{34}$ but otherwise
functional and aminoacylated.
\item Hypomodified tRNA decodes its cognate codons more slowly, in a manner dependent jointly
on A-site and P-site identity.
\item On transcripts with high initiation rate and clustered vulnerable codon pairs, the
additional dwell time drives queue formation past the threshold for collision.
\item Collisions engage collision-sensing ubiquitination and canonical no-go decay, degrading
the transcript.
\end{enumerate}

\subsection{Evidential status, link by link}

\paragraph{Links 3 and 5 are supported.} Real-time kinetic work shows that loss of wobble
modification lowers cognate-codon binding affinity, increases proofreading rejection, and
slows both post-hydrolysis rearrangement and translocation \cite{rodnina2017}. Ribosome- and
disome-profiling in $\mathrm{U}_{34}$-modification-null yeast shows stalling at specific
modification-dependent codon \emph{pairs}, triggering collisions recognized by the collision
sensor Hel2 and degradation through canonical no-go decay initiated by Cue2 cleavage
\cite{simms2017,nar2025}. These are the strongest foundations the hypothesis has.

\paragraph{Link 4 is plausible but unquantified.} The same work establishes that most
suboptimal codons are translated without detectable effect, and that only specific pairs
induce collisions \cite{nar2025}. The threshold is therefore real but narrow, and where human
transcripts sit relative to it is unknown.

\paragraph{Links 1 and 2 are the weakest, and at present unsupported.} I am aware of no
published demonstration of ASO-mediated steric blockade of pre-tRNA modification. The ASO
must win an occupancy race against modification enzymes acting on a folded, protein-bound,
co-transcriptionally processed substrate, at nuclear free-ASO concentrations that are low and
poorly characterized. More seriously, the one published instance of ASO targeting of a tRNA
reports a different outcome: no reduction in tRNA level, the effect attributed to steric
sponging that blocks interaction with synthetases, ribosomes and elongation factors, with the
downstream response running through ribotoxic stress and nonsense-mediated decay
\cite{orellana2026}.

This distinction is not a detail. \emph{Sequestration} removes functional tRNA and produces a
broad, codon-family-wide slowdown. \emph{Hypomodification} leaves tRNA present but slow at
specific codon pairs. Only the second supports the selectivity argument. The framework's
central claim rests on obtaining the phenotype the available evidence makes less likely, and
must be treated as conditional on that point being resolved experimentally.

\subsection{The selectivity premise}

The therapeutic proposition requires that vulnerable-pair burden separate transcripts whose
loss is desirable from transcripts whose loss is toxic. This premise is assumed, not
demonstrated, and there is a structural reason for concern: the substrate of the mechanism is
high translational flux, and high flux is a property of constitutively essential genes at
least as much as of oncogenes. Consistent with this, collision-triggered degradation in yeast
has been reported to affect abundant, essential metabolic enzymes \cite{simms2017}. Whether an
exploitable separation exists in the human transcriptome is an empirical question with a
computable answer; it is the first milestone below.

A mechanism acting through the shared decoding apparatus has no intrinsic tumour specificity:
any selectivity must derive either from codon-usage statistics or from delivery. Claims of a
safety margin based on preserved tRNA charging---for instance, that uncharged-tRNA-sensing
stress pathways such as GCN2 are thereby bypassed---are contingent on link 2 holding, and
should not be advanced until it does.

\section{Worked Illustration: Construction of a Case Study}

I deliberately present no case study as a result. To indicate what one would require,
consider a candidate oncogene. The analysis needs the canonical coding sequence; the
empirically determined set of vulnerable ordered codon pairs; the observed burden and
clustering statistics; a synonymous-shuffle null preserving the encoded protein and
genome-wide codon frequencies; and a comparison against length-, $\mathrm{GC}_3$- and
ribosome-density-matched essential genes. Only the last makes the comparison informative,
since enrichment relative to the genome average is largely a restatement of expression level.

The relevant tRNAs have to be tracked consistently. The
$\mathrm{mcm}^{5}\mathrm{U}_{34}$-modified set comprises $\mathrm{tRNA}^{\mathrm{Lys}}_{UUU}$,
$\mathrm{tRNA}^{\mathrm{Glu}}_{UUC}$ and $\mathrm{tRNA}^{\mathrm{Gln}}_{UUG}$, decoding AAA,
GAA and CAA respectively. Codons decoded by other families, including AGA
($\mathrm{tRNA}^{\mathrm{Arg}}$), fall outside this mechanism and must not be aggregated with
them; conflating families would inflate any apparent enrichment.

\section{Open Challenges and Pre-Registered Roadmap}
\label{sec:roadmap}

I state two analytical milestones with decision rules fixed in advance, including the
outcomes under which the programme narrows or terminates, followed by the minimal viable
experiment.

\subsection{Milestone 1: selectivity separation}
\label{sec:milestone1}

\paragraph{Design.} Construct the vulnerable-pair set $\mathcal{V}$ from published disome-seq
in $\mathrm{U}_{34}$-null backgrounds, with the enrichment threshold fixed before examining
human data, and with strict, permissive, and codon-family-only variants all reported. Draw
oncogenes from a curated source; draw the counter-set from genome-wide CRISPR
common-essentials rather than a loose housekeeping list. Construct a matched counter-set on
coding-sequence length, $\mathrm{GC}_3$, and ribosome density.

For a coding sequence $g$ of length $L_g$ codons, with indicator
$v_i = \mathbf{1}\!\left[(c_i, c_{i+1}) \in \mathcal{V}\right]$ over consecutive P/A pairs,
define the burden and the footprint-scale clustering statistic
\begin{align}
\rho_g &\;=\; \frac{1}{L_g - 1} \sum_{i=1}^{L_g-1} v_i,
\label{eq:statistics}\\
\kappa_g(w) &\;=\; \max_{i} \sum_{j=i}^{i+w-1} v_j, \notag
\end{align}
with window $w$ pre-specified at the ribosome-footprint scale and sensitivity reported over a
range of $w$. Against a synonymous-shuffle null preserving the encoded protein and
genome-wide codon frequencies, report
$z_g = \left(\kappa_g - \mu_g^{\mathrm{null}}\right) / \sigma_g^{\mathrm{null}}$.

\paragraph{Endpoint.} The area under the receiver operating characteristic curve (AUC), with
bootstrap confidence interval, discriminating oncogenes from the \emph{matched} counter-set,
reported alongside the partial AUC restricted to the high-specificity regime, since that is
the only regime relevant to a therapeutic window.

\paragraph{Decision rule.} Matched $\mathrm{AUC} \ge 0.75$, with confidence-interval lower
bound above $0.70$ and partial AUC clearly above chance: the selectivity premise survives, as
a necessary but not sufficient condition. $0.60 \le \mathrm{AUC} < 0.75$: separation too weak
to support a therapeutic claim; the programme is reframed as a quantitative model of
collision-mediated translational control, with the oncological application demoted to
discussion. $\mathrm{AUC} < 0.60$, or separation vanishing under matching: the premise
collapses into non-selective translation inhibition and is reported as such. Required
controls: a scrambled pair set of matched frequency must yield $\mathrm{AUC} \approx 0.5$; the
result must survive an independent oncogene list and restriction to a single tumour type's
expressed transcriptome.

\subsection{Milestone 2: reference invariance and non-redundancy}
\label{sec:milestone2}

\paragraph{Invariance.} Verify that the proposed perturbation satisfies the
multiplicative-on-fixed-graph condition of Section~\ref{sec:reference}; where it does not,
compute and report the bound \eqref{eq:bound} rather than suppressing the ambiguity.

\paragraph{Sanity battery.} Each check carries an explicit failure condition.
\begin{enumerate}
\item[(a)] \emph{Equilibrium limit.} At zero thermodynamic drive, $S \to 0$ and $K$ reduces to
the detailed-balance traffic, verified analytically. A residual $S \neq 0$ indicates a sign or
reference error.
\item[(b)] \emph{Linear response.} The response function must decompose into an entropic term
recovering the equilibrium fluctuation--dissipation relation and a frenetic correction that
vanishes as the drive tends to zero. A non-vanishing correction at equilibrium is a failure.
\item[(c)] \emph{Thermodynamic uncertainty.} For the collision count $N_{\mathrm{coll}}(T)$
with mean $J$ and entropy-production rate $\Sigma$ computed independently from the hydrolysis
flux, simulated statistics must satisfy
\begin{equation}
\frac{\mathrm{Var}\!\left[N_{\mathrm{coll}}(T)\right]}{J^{2}}
\;\ge\;
\frac{2 k_{B}}{\Sigma\, T},
\label{eq:tur}
\end{equation}
with the generalized bound substituted where the projected dynamics is non-Markovian.
Violation indicates an unaccounted drive absorbed into an effective rate.
\item[(d)] \emph{Vanishing-initiation limit.} The collision density $\Phi$ must approach zero
continuously as the initiation rate $\alpha \to 0$, with leading behaviour
$\Phi \sim \alpha^{2}$, collisions being pairwise. A non-zero intercept or linear scaling
indicates that the observable is not counting collisions.
\item[(e)] \emph{Finite-window consistency.} The attribution must be stable over observation
windows comparable to those a profiling experiment provides, the preceding checks being
asymptotic statements.
\end{enumerate}

\paragraph{Non-redundancy.} Exhibit at least one regime in which the decomposition predicts
collision statistics that a matched rate-level exclusion-process model predicts incorrectly,
or report that no such regime was found. This is the condition under which the physics is
load-bearing rather than descriptive; absent it, the correct conclusion is that the queueing
model suffices and the decomposition should be dropped.

\subsection{Milestone 3: minimal viable experiment}

The natural first experiment does not employ an ASO. Genetic depletion of $\mathrm{U}_{34}$
modification in a system where the collision phenotype is already established, combined with
matched ribosome- and disome-profiling across a titration of initiation rate, tests the
physics without confounding it with unresolved delivery chemistry. The model must predict,
before unblinding, which codon pairs become collision-prone and the non-linear form of the
dependence on $\alpha$. A reporter panel with engineered codon-pair spacing tests the
specifically queueing-theoretic prediction; a collision-sensor-null background separates
collision-triggered decay from slow-down alone. Only if these succeed does the ASO arm become
worth attempting, and it should then be posed first as a question---does the ASO produce
hypomodification or sequestration?---rather than as a therapeutic test.

\subsection{Further open problems}

Formalization of the trap, in the sense of Section~1.2, requires a fitness functional and an
explicit space of accessible compensations. The vulnerable-pair set is derived in yeast and
applied to human sequence, an extrapolation that is currently an assumption. The validity of
the scalar reaction coordinate remains to be established. Selectivity derived from codon
statistics must be distinguished from selectivity derived from delivery. Finally, the
resistance question stands in its own right: reduced initiation rate may constitute a cheap
escape, since lowering $\alpha$ dissolves queues without any change to codon usage.

\section{Concluding Remarks}

The framework's appeal is that it treats translation as a driven flow, not a set of
concentrations, and that collision-triggered decay supplies a genuine non-linearity for a
timing perturbation to act on. Its weakness is that both decisive claims---selectivity in the
human transcriptome, and predictive content beyond existing traffic models---are at present
assumptions. Both are testable now, cheaply, before any wet-laboratory commitment. Running
them, and reporting the outcome whichever way it falls, is the programme's actual next step.
What is stated here is a specification of those tests, not a claim to have passed them.


{\small
\begin{thebibliography}{99}
\setlength{\itemsep}{1pt}
\setlength{\parsep}{0pt}
\setlength{\parskip}{0pt}

\bibitem{maes2019}
C.~Maes,
\newblock Frenesy: time-symmetric dynamical activity in nonequilibria,
\newblock \emph{Physics Reports} (2020); arXiv:1904.10485.

\bibitem{basumaes2015}
U.~Basu and C.~Maes,
\newblock Nonequilibrium response and frenesy,
\newblock \emph{J.~Phys.: Conf.~Ser.} (2015); arXiv:1507.01228.

\bibitem{segal2026}
S.~Segal,
\newblock Channel selection at identically vanishing dissipation difference: isolating the
frenetic sector of the overdamped path measure,
\newblock arXiv:2608.00041 [physics.chem-ph] (2026).

\bibitem{lecomte2012}
V.~Lecomte, J.~P.~Garrahan, and F.~van Wijland,
\newblock Inactive dynamical phase of a symmetric exclusion process on a ring,
\newblock \emph{Journal of Physics A: Mathematical and Theoretical} \textbf{45}, 175001
(2012).

\bibitem{gjw2007}
J.~P.~Garrahan, R.~L.~Jack, V.~Lecomte, E.~Pitard, K.~van Duijvendijk, and F.~van Wijland,
\newblock Dynamical first-order phase transition in kinetically constrained models of glasses,
\newblock \emph{Physical Review Letters} \textbf{98}, 195702 (2007).

\bibitem{sahamukherji2015}
B.~Saha and S.~Mukherji,
\newblock Entropy production and large deviation function for systems with microscopically
irreversible transitions,
\newblock arXiv:1511.02587 (2015).

\bibitem{quasteltsai2021}
J.~Quastel and L.-C.~Tsai,
\newblock Hydrodynamic large deviations of TASEP,
\newblock \emph{Communications on Pure and Applied Mathematics}; arXiv:2104.04444 (2021).

\bibitem{reuveni2011}
S.~Reuveni, I.~Meilijson, M.~Kupiec, E.~Ruppin, and T.~Tuller,
\newblock Genome-scale analysis of translation elongation with a ribosome flow model,
\newblock \emph{PLoS Computational Biology} \textbf{7}, e1002127 (2011).

\bibitem{subramaniam2017}
A.~R.~Subramaniam et~al.,
\newblock Kinetic modeling predicts a stimulatory role for ribosome collisions at elongation
stall sites in bacteria,
\newblock \emph{eLife} \textbf{6}, e23629 (2017).

\bibitem{simms2017}
C.~L.~Simms, L.~L.~Yan, and H.~S.~Zaher,
\newblock Ribosome collision is critical for quality control during no-go decay,
\newblock \emph{Molecular Cell} (2017).

\bibitem{nar2025}
Suboptimal codon pairs trigger ribosome collisions and cellular quality control responses in
tRNA modification mutants,
\newblock \emph{Nucleic Acids Research} \textbf{53}, gkaf1311 (2025).

\bibitem{rodnina2017}
Thio-modification of tRNA at the wobble position as regulator of the kinetics of decoding and
translocation on the ribosome,
\newblock \emph{Journal of the American Chemical Society} \textbf{139} (2017).

\bibitem{orellana2026}
E.~A.~Orellana et~al.,
\newblock Targeting tRNA-Arg-TCT-4-1 suppresses cancer cell growth and tumorigenesis,
\newblock bioRxiv (2026).

\bibitem{alkbh8}
ALKBH8-mediated codon-specific translation promotes colorectal tumorigenesis,
\newblock \emph{Nature Communications} \textbf{16} (2025).

\end{thebibliography}
}
\end{document}